# Stretching, Twisting and Supercoiling in Short, Single DNA Molecules

Pui-Man Lam[*]

Physics Department, Southern University

Baton Rouge, Louisiana 70813

and

Yi Zhen[+]

Department of Natural Sciences, Southern University

6400 Press Drive

New Orleans, Louisiana 70126

**Absstract:**

We had combined the Neukirch-Marko model that describes the extension, torque and supercoiling in single, stretched and twisted DNA of infinite contour length, with a form of the free energy suggested by Sinha and Samuels to describe short DNA, with contour length only a few times the persistence length. We find that the free energy of the stretched but untwisted DNA, is significantly modified from its infinitely length value and this in turn modifies significantly the torque and supercoiling. We show that this is consistent with short DNA being more flexible than infinitely long DNA. We hope our results will stimulate experimental investigation of torque and supercoiling in short DNA.

PACS numbers: 87.14.gk, 82.35.Rs, 87.15.A-, 87.15.La

[*]puiman_lam@subr.edu;  [+]yzhen@suno.edu

## I. Introduction

Changes in the supercoiling state of the DNA double helix under tension and torque are ubiquitous in cellular level and affect virtually all genomic processes [1]. Proteins moving along the helical path of the DNA, for example, generate torsional stress, which produces the over- or under-winding of the DNA around its axis and coiling the duplex axis around itself. Locally destabilized or strand separation under stress is required for the initiation of both transcription and protein binding [2-4]. Twisted state of DNA can induce global changes in the conformation of the genome which brings distant DNA sequences together to facilitate specific localization of replication initiation [5,6]. Genomic DNA is organized in topological domains of 10-100 kilobases (kb). Within these regions, torsion can rapidly transmit to allow for long-range communication between distant genomic locations [7]. To understand the cellular processes that are affected by changes in the supercoiling state, it is essential to elucidate the mechanisms of the behavior of DNA under torsion and the torque.

With development of the techniques for observing and manipulating single biological molecule, experimental methods such as optical and magnetic tweezers, microfibers and atomic force microscopy are available for direct measurement and visualization of the displacement and mechanical response of a single molecule under stretching force and torsional stress [8-17]. Experiments on stretched, supercoiled DNA primarily measure the extension of the molecule as a function of its twisting and change in double helix linking number Lk [18]. Theoretical explanation of these experimental results usually are deduced from traditional thermodynamic treatments starting with construction of free energy characterized by a bending rigidity A and a torsional rigidity [19-23]. In order to theoretically describe the experimental measurement of Mosconi et al.[18] of the buckling torque of DNA as a function of force in various salt concentration by a magnetic trap system, a heuristic model was made by Marko [24] and Neukirch and Marko [23] who suggested an analytical model to describe the behavior of the mixture of extended and supercoiled DNA molecule under twist. Experiments on stretched, supercoiled DNA primarily measure the extension of the molecule as a function of its twisting and change in double helix linking number Lk [18]. Theoretical explanation of these experimental results usually are deduced from traditional thermodynamic treatments starting with construction of free energy characterized by a bending rigidity  and a torsional rigidity.

The semi-flexible polymer model provides a quantitative starting point for theories describing these type of experiments [19-23]. In ref. [24] Marko has suggested a heuristic model to describe the behavior of a stretched DNA under twist. In this model, the twisted, stretched molecule is partitioned into an unstretched, plectonemic supercoil phase with torsional stiff $P$ and a stretched and twisted DNA molecule with bending rigidity $A$ and effective torsional stiffness $C_s$. The torsional stiffness $P$ of the plectonemic DNA is unknown and can only be estimated. The predictions of the model are in qualitative agreement with experiment [18]. This is not surprising since a description of the plectonemic phase with a single force independent torsional stiff $P$ is an oversimplification that does not take into account for example the variation of plectonemic radius with force due to entropic repulsion [25].

More recently, Neukirch and Marko [23] proposed an improved model that describes the stretched, twisted DNA molecule by taking into account the variation of the plectonemic radius with force explicitly. In this way the theory is completely parameter-free. In the limit of high force, closed-form asymptotic solutions for the supercoiling radius, extension and torque of the molecule can be obtained. These asymptotic solutions already give rather reasonable agreement with experiment. We have recently shown [26] that much of the discrepancy of this theory with experiment is due to the use of an approximate form of the free energy for the extended phase of the molecule, which is stretched but untwisted and that the use of a more accurate form of the free energy significantly improves the agreement with experiment. However, the theory of Neukirch and Marko[23] and the experimental results of Mosconi et al [18] refer to an extremely long DNA. In the experiment of Mosconi et al, the total length of the DNA is $5.4\,mm$, or more than 100 times the persistence length of $50\,nm$.

Sinha [27] and Samuel had shown that for short DNA, with contour length of the order of the persistence length, the free energy is modified compared to the case of infinite contour length. In this paper we calculate the modified free energy for short DNA and use it in the Neukirch-Marko theory to calculate the slope of the average extension with respect to linking number and the torque in a twisted DNA. We find that both quantities are significantly modified compared to the infinite contour length case. We hope our results will stimulate experimental investigations of these quantities in short DNA.

In section II we will summarize the Neukirch-Marko model. In section III we will give the form of the free energy in the case of infinite length DNA. In section IV we will show how the free energy of short DNA is modified from its infinite length case and the result of this modified free energy in calculating the slope and torque, using the Neukirch-Marko model. Section IV is the conclusion.

## II. The Neukirch-Marko Model

We will quote the main results of ref. [23] which has been summarized in ref. [26]. Fig. 1 is an illustration of a supercoiled DNA molecule under force $f$ and torque. The total length $L$ of the molecule is partitioned between the two "phases": (i) a plectonemic phase of length $l$, where the filament has bending rigidity $A$ and torsional rigidity $C$ and adopts a superhelical shape of radius $r$ and angle $a$, and (ii) an extended wormlike-chain phase of length $L-l$.

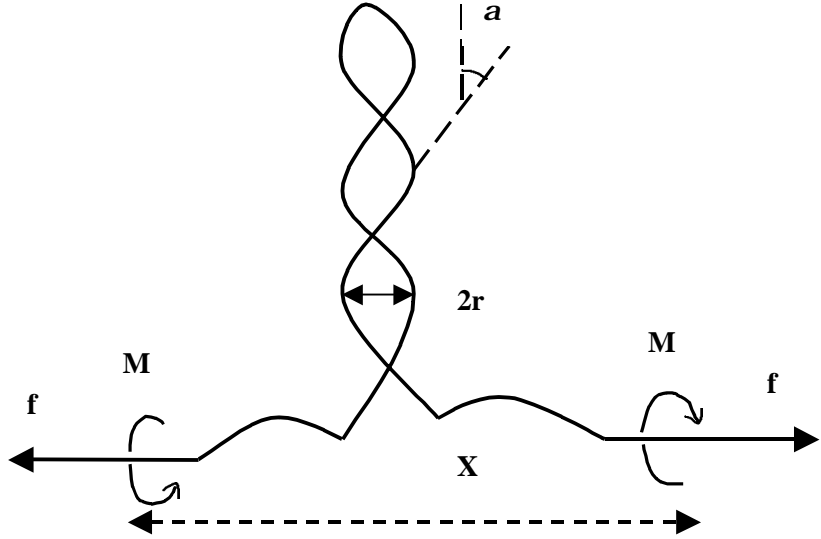

Figure 1. Supercoiled DNA under force and torque. Molecule length is partitioned between two phases: an extended phase and a plectonemic phase where strong self-interaction occurs.

When the torque is strong enough so that a plectonemic phase exists, it is shown in ref. [23] that the torque $M$ as a function of the force $f$ is given by the formula

$$M \approx [(32/27)A]^{1/4} g(f)^{3/4} / \sqrt{k_D} \sqrt{\log\left[\sqrt{9p/8}n^2 L_B k_B T / g(f)\right]}(1+a^2) \quad (1)$$

Here $g(f)$ is the per unit length free energy. $n$ is the effective charge given by the formula

$$n = \frac{1}{b} \frac{1}{g(L_B b, k_D a)} \frac{1}{k_D a K_1(k_D a)} \quad (2)$$

where $b = 0.17$ nm is half of the 0.34 nm spacing of successive base pairs along DNA; $a = 1$ nm is the radius of the cylindrical double helix. $L_B = 0.7$ nm is the Bjerrum length in water, $k_D^{-1}$ the Debye length, and $K_n(x)$ the nth modified Bessel function of the second kind [29, 30]. From Table III of ref [29], the parameter $g$ is computed to be $g = (1.64, 1.44, 1.27, 1.14)$ at salt concentrations (50, 100, 200, 500) mM and for $T = 296.5$ K.

Another quantity of experimental interest is the slope of the average extension $q = -2p\partial M / \partial f$. According to ref. [23], this quantity is given by the formula

$$q = \left(\frac{6A}{k_D^2 g(f)}\right)^{1/4} \sqrt{\log\left[\sqrt{9p/8}n^2 L_B k_B T / g(f)\right]} g'(f)(1 + 2a^2/3) \qquad (3)$$

In the model the free energy of the plectonemic phase is that of two straight charged cylinders with a center axis separated by a distance of $2r$, and the Debye–Hückel approximation is used in the resulting Poisson-Boltzmann equation.

The free energy of the extended phase is described in terms of the free energy per unit length of the untwisted molecule $g(f)$ [23], plus a twist energy using a twist modulus that includes effects of writhing fluctuations: $C_s(f) = C\left[1 - (C/4A)k_B T / \sqrt{Af}\right]$ [28]. For simplicity, the form $C_s = C$ was used. Experimental values of $A/k_B T = 46, 47, 44, 45$ nm at 50, 100, 200, and 500 nM salt and $C/k_B T = 94$ nm are used in the calculation.

It should be pointed out that the Neukirch-Marko model does not contain any other parameters except those determined by experiment. Also both the torque $M(f)$ and the slope $q(f)$ depend explicitly on the per unit length free energy of the untwisted molecule $g(f)$. This makes the theory applicable to various situations when the quantity $g(f)$ is known.

## III. Free Energy for Infinitely Long DNA

The calculation of the slope and the torque using Eqns. (3) and (1) depend on the free energy of an untwisted but stretched DNA as input. The force-extension curve in the worm like chain (WLC) model is given by the widely used interpolation formula [22]

$$f = \frac{(k_B T)}{L_p}\left[\frac{X}{L} + \frac{1}{4}\left(1 - \frac{X}{L}\right)^{-2} - \frac{1}{4}\right], \qquad (4)$$

where $L_p$ here is the persistence length, related to the bending rigidity $A$, by $A = k_B T L_p$, and $X$ is the extension. The negative of the free energy per unit length $g(f)$ is obtained by a Legendre transform

$$Lg(f) = fX - W(X) \qquad (5)$$

where

$$W(X) = \int_0^X dX' f(X') \qquad (6)$$

is the work done in extending the polymer. The functions $g$ and $W$ depend also on the persistence length $L_p$. From Eqn. (4) the extension $X$ is an implicit function of the force $f$. Since the extension is a single-valued, monotonic increasing function of $f$, we can define the inverse function $X_f(f)$ which give the extension $X$ as a function of the force $f$. Even though this function cannot be obtained analytically, it can be calculated numerically to high accuracy. Substituting Eqn. (4) into Eqn. (6), the function $W$ can be calculated analytically:

$$W(X(f)) = L\frac{k_B T}{4L_p}\left[\frac{X_f(f)}{L}\left(2\frac{X_f(f)}{L}-1\right)+\left(1-\frac{X_f(f)}{L}\right)^{-1}\right] \quad (7)$$

From Eq. (5), the negative of the free energy per unit length is given as a function of the force $f$ by

$$g(f) = \frac{1}{L}fX_f(f) - \frac{k_B T}{4L_p}\left[\frac{X_f(f)}{L}\left(2\frac{X_f(f)}{L}-1\right)+\left(1-\frac{X_f(f)}{L}\right)^{-1}\right] \quad (8)$$

We had used [26] this form of the unit length free energy in Eqs. (1) and (3) to calculate the torque $M$ and slope $q$.

From Eq. (3), in order to calculate the slope $q$, the derivative of $g$ with respect to $f$ is needed. From Eqs. (5) and (6), this is given by

$$Lg'(f) = X_f(f) \quad (9)$$

## IV. Free Energy for Short DNA and Resulting Slope and Torque

Statistical mechanics of a single polymer molecule is dominated by fluctuations because it is a system of finite size. It is only in the thermodynamic limit of extremely long polymers that these fluctuations about the mean die out. However experimental interest is *not confined to very long polymers*. For example, experiments on Actin [32] deal with polymers of length L = 30ì m, which is only about twice the measured persistence length of Lp = 16.7ì m. There is also theoretical work [33] on extremely short polymers.

Due to the dominance of fluctuations, the experimentally measured mean values for a semi-flexible polymer crucially depend on the precise choice of the ensemble. For instance, one gets qualitatively distinct features in force-extension curves depending on whether the force or the extension is held constant in an experimental setup [34, 35].

In statistical mechanics, an isometric setup would be described by the Helmholtz free energy, whereas an isotensional setup would be described by the Gibbs free energy [36]. In the thermodynamic limit these two descriptions agree, but semiflexible polymers (those with contour lengths comparable to their persistence lengths) are *not* at the thermodynamic limit. Experimentally, both isometric and isotensional ensembles are realizable. It turns out, that an experiment in which the ends of a polymer molecule are fixed (isometric) and the force fluctuates yields a different result from one in which the force between the ends is held fixed (isotensional) and the end-to-end distance fluctuates [32, 37, 38]. This difference can be traced to large fluctuations about the mean value of the force or the extension, depending on the experimental setup. These fluctuations vanish only in the thermodynamic limit of very long polymers.

Sinha and Samuel [27] had studied DNA under tension. They found that short DNA under constant tension is described by the Gibbs ensemble, with unit length free energy $g(f)$ given by

$$\exp(L\tilde{g}(\tilde{f})) = \int_0^1 du \exp\left(-\frac{L[\tilde{w}(u) - \tilde{f}u]}{L_p}\right) \quad (10)$$

where the dimensionless quantities $u$, $\tilde{f}$, $\tilde{g}$ and $\tilde{w}$ are given by

$$u = \frac{X}{L}, \quad \tilde{f} \equiv \frac{fL_p}{k_B T}, \quad \tilde{g}(\tilde{f}) \equiv \frac{g(f)}{k_B T}, \quad \tilde{w}(u) \equiv \frac{L_p W(X)}{L k_B T}. \quad (11)$$

In the limit of an infinite long chain, $L \to \infty$, the integral can be replaced by the integrand evaluated at the value of $u$ that will give the largest value in the exponent. This is given by the condition

$$\frac{d}{du}[\tilde{w}(u) - \tilde{f}u] = 0$$

This will yield the usual Legendre transform Eqn. (5) of an infinitely long DNA

$$Lg(f) = -W(X) + fX \quad \text{with} \quad W(X) = \int_0^X f(X')dX'.$$

Using Eqn. (7) in the form

$$\tilde{w}(u) = u(2u - 1) + (1 - u)^{-1},$$

the unit length free energy for a short DNA from Eqn. (10), is given by

$$\tilde{g}(\tilde{f}) = \frac{1}{L} \ln\left\{\int_0^1 du \exp\left(-\frac{L}{4L_p}[u(2u-1) + (1-u)^{-1}] + \frac{L}{L_p}\tilde{f}u\right)\right\}. \quad (12)$$

In Figure 2 we show the free energy for short DNA with contour length $L = 6L_p$ calculated using equation (12), compared with that of infinitely long DNA calculated using equation (8), for various salt concentrations. We find that different salt concentrations do not produce distinguishable results for the unit length free energy in this figure, both for short and infinitely long DNA. For short DNA the contour length used is $L = 6L_p$. But for contour lengths from $L = 6L_p$ to $L = 10L_p$ we find similar results for the unit length free energy.

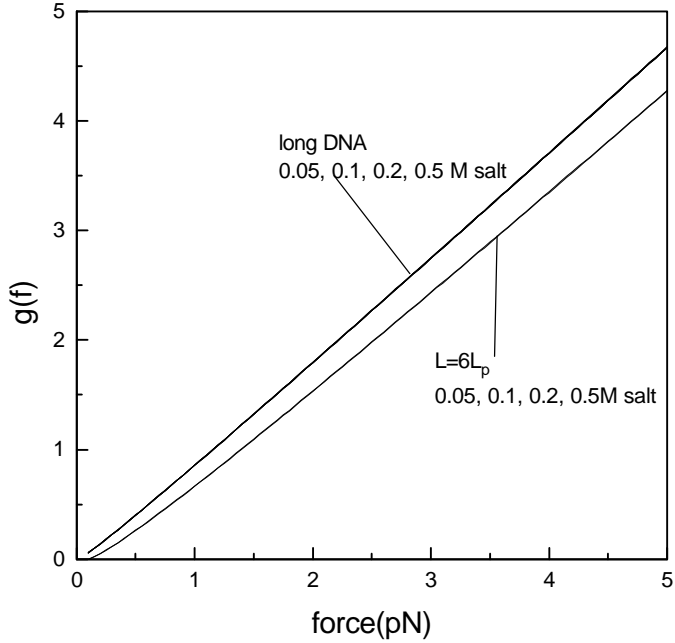

Figure 2: Free energy g(f) versus force for infinitely long DNA and for short DNA with contour length $L = 6L_p$. Different salt concentrations produce indistinguishable results.

We have used this form of the unit length free energy for short DNA in equations (3) and (1) to calculate the slope and torque. In Figure 3 we show the results for the slope versus force, for short DNA with contour length $L = 6L_p$, and for infinitely long DNA, at various salt concentrations. The results for the slope are slightly different for the two cases. In Figure 4 we show the results for the torque versus force, for short DNA with contour length $L = 6L_p$, and for infinitely long DNA, at various salt concentrations. The results for the torque are noticeably different for the two cases. In fact, the torque for the short DNA is always smaller than that of the infinitely long DNA. This can be understood from equation (1) for the torque. The torque is essentially proportional to $g(f)^{3/4}$. From Figure 2, we know that the free energy $g(f)$ for short DNA is always smaller than that of the infinitely long DNA and that leads to the smaller value of the torque for short DNA.

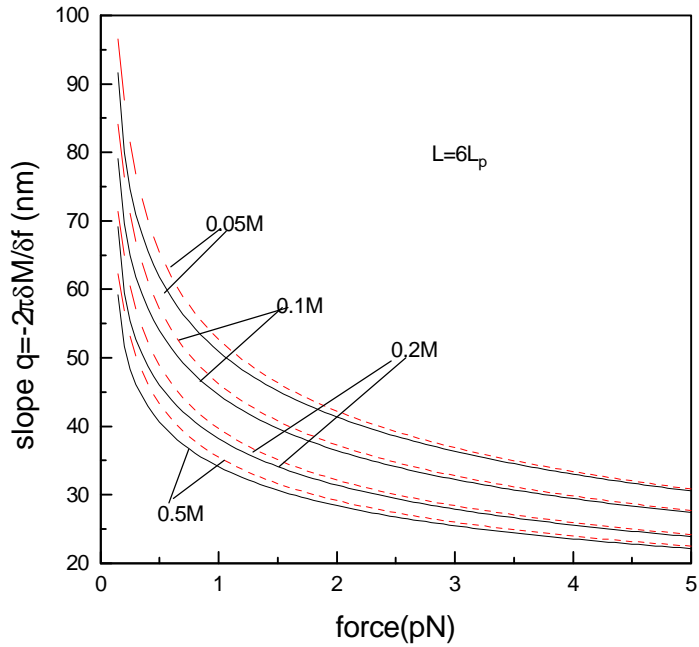

Fig. 3. The slopes $q = -2p\partial M/\partial f$ of the torque, as a function of the applied force, for 0.05, 0.1, 0.2 and 0.5 M salt (top to bottom). Full lines are our theoretical results for short DNA with contour length $L = 6L_p$. Dashed lines are theoretical results for infinitely long DNA.

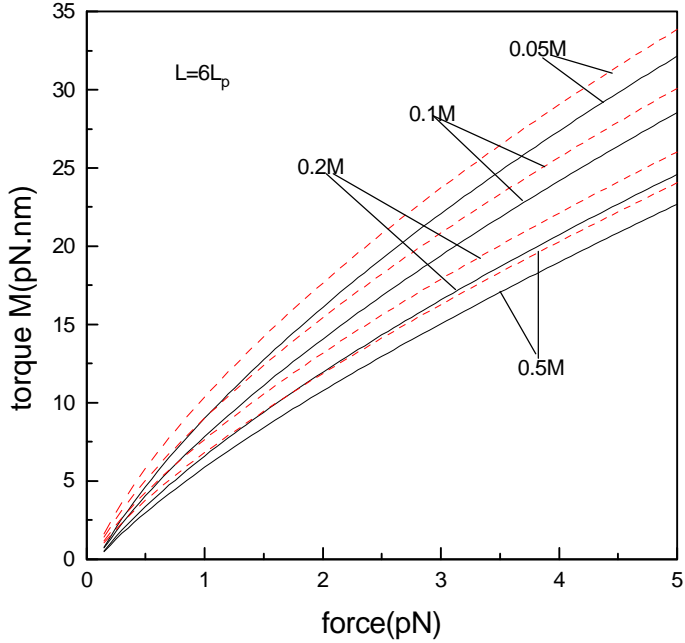

Fig. 4. Theoretical torque as a function of the applied force, for 0.05, 0.1, 0.2 and 0.5 M salt (top to bottom). Full lines are our theoretical results for short DNA with contour length $L = 6L_p$. Dashed lines are theoretical results for infinitely long DNA.

**V. Conclusion**

The central idea of this paper is that, when the contour length DNA is few times of persistence length, the construction of free energy cannot be theoretically treated like that of long DNA. We proposed analytical description of free energy of short DNA under tension. The free energy of short DNA is constructed using Laplace transform instead of Legendre transform. We applied this modified free energy for short DNA in the Neukirch-Marko theory to calculate the slope of the average extension with respect to linking number and the torque in a twisted DNA. We have shown that both quantities, especially the torque, are significantly modified compared to the infinite contour length case. The distinction between the short and infinitely long DNA can be traced to the fact that in the former case, there exists large fluctuations about the mean value of the force or the extension, depending on the experimental setup. An isometric setup (an experiment in which the ends of a polymer molecule are fixed and the force fluctuates) would be described by the Helmholtz free energy and a isotensional set up ( an experiment in which the force between the ends is held fixed and the end-to-end distance fluctuates) would be described by the Gibbs free energy. The partition function in Gibbs ensemble is the Laplace transform of that in Helmholtz ensemble. In the thermodynamic limit of long polymer, the Laplace transform integral is dominated by the saddle point value and therefore Gibbs free energy and Helmholtz free energy are related by a Legendre transform.

It is noted that due to the dominance of fluctuations, the experimentally measured mean values for a semi-flexible polymer crucially depend on the precise choice of the ensemble. In the thermodynamic limit, Gibbs free energy and Helmholtz free energy descriptions agree, but semi-flexible polymers (those with contour lengths comparable to their persistence lengths) are *not* at the thermodynamic limit. These fluctuations vanish only in the thermodynamic limit of very long polymers. In Figure 2 we show the free energy for short DNA with contour length $L = 6L_p$, compared with that of infinitely long DNA, for various salt concentrations. It can be seen that different salt concentrations do not produce distinguishable results for the unit length free energy in this figure, both for short and infinitely long DNA. For short DNA the contour length used is $L = 6L_p$. But for contour lengths from $L = 6L_p$ to $L = 10L_p$ we find similar results for the unit length free energy. The results for the slope are slightly different for the two cases. In Figure 4 we show the results for the torque versus force, for short DNA with contour length $L = 6L_p$, and for infinitely long DNA, at various salt concentrations. The results for the torque are noticeably different for the two cases. In fact, the torque for the short DNA is always smaller than that of the infinitely long DNA. This can be understood from equation (1) for the torque. The torque is essentially proportional to $g(f)^{3/4}$. From Figure 2, we know that the per unit length free energy $g(f)$ for short DNA is always smaller than that of the infinitely long DNA and that leads to the smaller value of the torque for short DNA.

Why do the results differ apparently for long and short DNAs? Some earlier experiments suggest that short DNAs have higher flexibility than long ones [39-41], while some recent so-called high-resolution experiments suggest that a DNA at short length scale can still be described by a worm-like chain model with parameters for long DNA [42,43], and the higher flexibility of short DNAs has been proposed to be attributed to the end effect based on all-atom MD simulations [44], or to the correlated local bending based on a mesoscopic model [45]. Our result that the per unit length free energy $g(f)$ is smaller for short DNA is actually consistent with short DNA being more flexible than long DNA. The reason is the following. Shorter DNA being more flexible means its persistence length $L_p$ is smaller for short DNA. From Eqn. (4), the force versus per unit length extension $(X/L)$ curve for short DNA is above that of long DNA. This implies that the per unit length extension $(X/L)$ versus force $f$ curve for short DNA is below that of long DNA, i.e. $(X/L)_{short} < (X/L)_{long}$. From Eqn. (8), one can see that the per unit length free energy $g(f)$ is smaller for short DNA.

One can ask the question if plectonemes can actually form in short DNA, since the bending energy cost may be too high to be balanced by the twist energy. To answer this question we will calculate the plectoneme radius and show that the results are physically meaningful. It is shown in reference [23] that solutions of the Neukirch-Marko model with finite plectonme length 0<l<L exist as soon as the linking number $\Delta Lk$ is large enough. Therefore it remains to show that the plectoneme radius calculated is physically meaningful. The plectoneme radius r is given by eqn. (14) in ref.[23]. We have used this formula to calculate the plectoneme radius as a function of the applied force for different salt concentrations, for DNA contour length L=6Lp. Our result is shown in Figure 5. Except for large salt concentration and large force, the radius is larger than 2 nm and is physically meaningful because it is larger than twice the DNA radius of about 1 nm.

Finally, our results call for experimental investigations of short DNA that might be explored in measurement of the extension of supercoiled short DNA as a function of its twisting and change in double helix linking number Lk. These results can be compared with existing results of very long DNA.

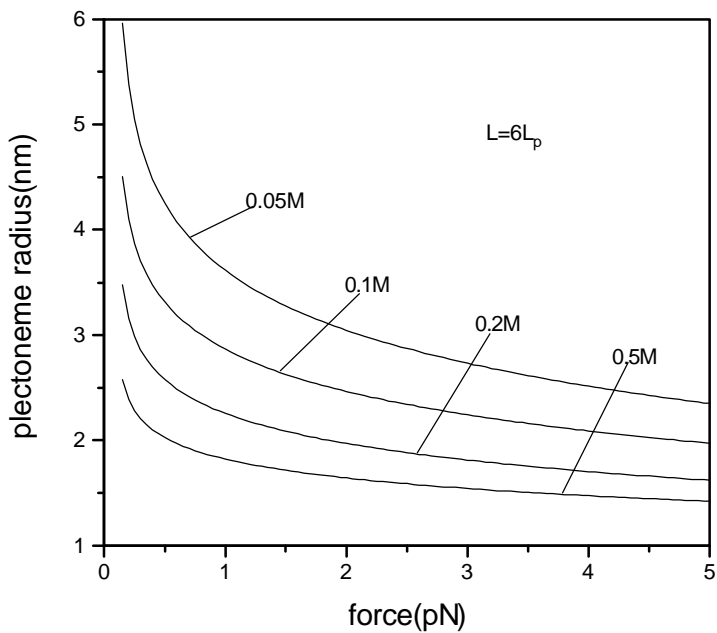

Figure 5: Plectoneme radius as function of force for 0.05, 0.1, 0.2 and 0.5 M salt, for short DNA with contour length $L = 6L_p$.

,